\documentclass[aps,prl,twocolumn,showpacs,superscriptaddress]{revtex4}
\usepackage[dvips]{graphicx}
\usepackage{amssymb}
\begin{document}

\title{Synchronization in Small--world Systems}

\author{Mauricio Barahona}~\altaffiliation{Present address:
Dept.\ of Bioengineering, Imperial College, London SW7 2BX,
UK.}~\affiliation{Control and Dynamical Systems, California
Institute of Technology, Pasadena, CA 91125}

\author{Louis M.\ Pecora}
\affiliation{Naval Research Laboratory, Code 6340,
 Washington, DC 20375}

\date{\today}

\begin{abstract}
We quantify the dynamical implications of the small-world
phenomenon. We consider the generic synchronization of oscillator
networks of arbitrary topology, and link the linear stability of
the synchronous state to an algebraic condition of the Laplacian
of the graph. We show numerically that the addition of random
shortcuts produces improved network synchronizability. Further, we
use a perturbation analysis to place the synchronization threshold
in relation to the boundaries of the small-world region. Our
results also show that small-worlds synchronize as efficiently as
random graphs and hypercubes, and more so than standard
constructive graphs.
\end{abstract}

\pacs{}

\maketitle

Recently, Watts and Strogatz~\cite{refWS} showed that the addition
of a few long-range shortcuts to an otherwise locally connected
lattice (the "pristine world") produces a sharp reduction of the
average distance between arbitrary nodes. The ensuing semi-random
lattice was denoted a \textit{small-world} (SW) because the sudden
appearance of short paths occurs early on, while the system is
still relatively localized. This concept has wide appeal: the SW
property seems to be a quantifiable characteristic of many
real-world
structures~\cite{refWS,refStevelong,refDuncanbook,refbiochem},
both human generated (social networks, WWW, power grid), or of
biological origin (neural and biochemical networks).

A spur of ongoing research~\cite{refStevelong} has concentrated on
static and combinatoric
properties~\cite{refBA,refBW,refNMW,refNW,refMM,refKAS} of a
tractable SW model~\cite{refWS,refRM}. Monasson~\cite{refRM}
considered the SW effect on the distribution of eigenvalues of the
connectivity matrix (the graph Laplacian) which specifies the
coupling between nodes---a relevant topic for polymer
networks~\cite{refpolymer}. However, despite their central role in
real-world networks, there are fewer studies of dynamical
processes taking place on SW lattices. Among those, automata
epidemics simulations~\cite{refepidemics} and Web-browsing
studies~\cite{refkleinberg} have revealed the importance of
shortcuts. Numerical work on synchronization of Kuramoto
oscillators~\cite{refDuncanbook}, discrete maps~\cite{refGH} and
Hodgkin-Huxley neurons~\cite{refLHC} has shown improved SW
synchronizability, as intuitively expected. However, these
numerical examples are not generic, and fail to provide insight
into how the SW property influences the dynamics.

In this paper, we explicitly link the SW addition of random
shortcuts to the synchronization of networks of coupled dynamical
systems. This is an example of dynamics {\it on}
networks---leaving aside the distinct problem of evolution {\it
of} networks here. By using a generic synchronization
formulation~\cite{refPC,refFJM} to factor out the connectivity of
the network, we identify the synchronization threshold with an
algebraic condition of the graph Laplacian. Through numerics and
analysis, we quantify how the SW scheme improves the
synchronizability of the pristine world, mainly as a result of the
steep increase of the first-non-zero eigenvalue (FNZE). The
synchronization threshold is found to lie in the SW
region~\cite{refepidemics,refDuncanbook}, but does not coincide
with its onset---it can in fact be linked to the effective
randomization that ends SW. Within this framework, we show that
the synchronization efficiency of semi-random SW networks is
higher than standard deterministic graphs, and comparable to both
fully random and ideal constructive graphs.

\begin{figure}
\vspace*{-.4in}
\includegraphics[width=3.5in]{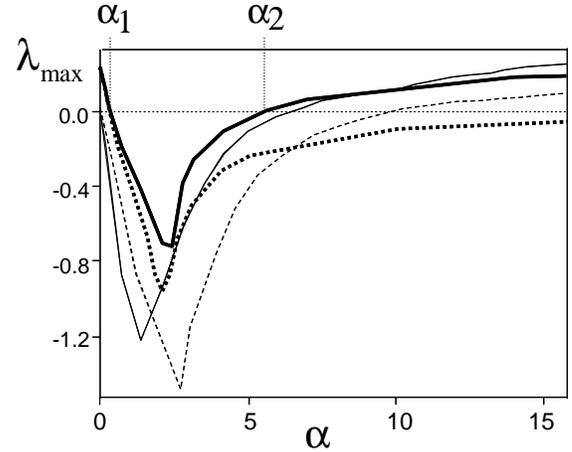}
\vspace*{-2.2in} \caption{Four typical master stability functions
(scaled for clearer visualization) for R\"ossler systems: chaotic
(bold) and periodic (regular lines); with $y$-coupling (dashed)
and $x$-coupling (solid lines). Here we consider the $x$-coupled
chaotic case (solid bold) with a negative well between
$(\alpha_1,\alpha_2)$.
\label{fig:masterfunction}}
\end{figure}

Consider $n$ identical dynamical systems (placed at the nodes of a
graph) that are linearly and symmetrically coupled (as represented
by the edges of the undirected graph) with global coupling
strength $\sigma$. The topology of the graph can be encoded in the
Laplacian matrix $G$, a symmetric matrix with zero row-sum and
real spectrum $\{\theta_k\}, \; k=0,1,\ldots, n-1$. A general
linear stability criterion for the synchronized state of the
system~\cite{refPC,refFJM} is given by the negativity of the
master stability function $\lambda_{\rm{max}} (\sigma \, \theta_k)
< 0,\; \forall k$. This $\lambda_{\rm{max}}$ is a characteristic
of the particular dynamics at the nodes but, crucially, a large
class of oscillatory systems (chaotic, periodic and quasiperiodic)
have master stability functions with generic
features~\cite{refFJM}. In particular, for several chaotic systems
$\lambda_{\rm{max}}$ has a single deep well, as depicted in
Fig.~\ref{fig:masterfunction}. (We remark that this analysis is
quite general: it can be extended~\cite{refFJM} to eliminate the
zero row-sum constraint, and to comprise nonlinear coupling and
more general synchronization criteria.) Stability is thus ensured
by tuning the coupling $\sigma$ to try and place the entire
spectrum of transverse eigenvalues (times $\sigma$) in the deep,
stable region: $\sigma \, \theta_k \in (\alpha_1,\alpha_2)$. This
leads to an algebraic condition for the existence of a linearly
stable synchronous state: a network is synchronizable if
\begin{equation}
\theta_{\rm{max}}/\theta_1 < \alpha_2/\alpha_1 \equiv \beta,
\label{eq:critcond}
\end{equation}
where $\theta_1$ is the FNZE and $\theta_{\rm{max}}$ is the
maximum eigenvalue of the Laplacian $G$. The figure of merit
($\beta$) ranges from 5 to 100 for a variety of oscillators (e.g.,
Lorenz, R\"ossler, double scroll).

Small-worlds are generated from a pristine world: a $k$-cycle of
$n$ nodes and range $k$, each node coupled to its $2 \, k$ nearest
neighbors for a total of $n k$ edges~\cite{refRM}. The Laplacian
of this graph $G^0$ is a banded circulant matrix with non-zero
elements on the main diagonal and the $2 \,k$ adjacent diagonals:
$G^0_{ii}=2 k$ and $G^0_{ij}=G^0_{ji}=-1$ with $(i+1)\bmod n \leq
j \leq (i+k) \bmod n$, and $1 \leq i \leq n$. The SW scheme dopes
the pristine world by adding $n s$ edges picked at random from the
$n (n-2k-1)/2$ remaining pairs. Each new edge between nodes $l$
and $ m$ adds off-diagonal $\Delta G_{lm}=\Delta G_{ml}=-1$ and
on-diagonal $\Delta G_{ll}=\Delta G_{mm}=1$ contributions to the
Laplacian, thus preserving the null row-sum and the bidirectional
coupling. The average number of shortcuts per node ($s$) is
related to other measures of randomness ($p$ and $q$) used
previously~\cite{refWS,refRM}: $s \equiv kp \equiv q
(n-2k-1)/(2n)$.

\begin{figure}
\includegraphics[width=3.3in]{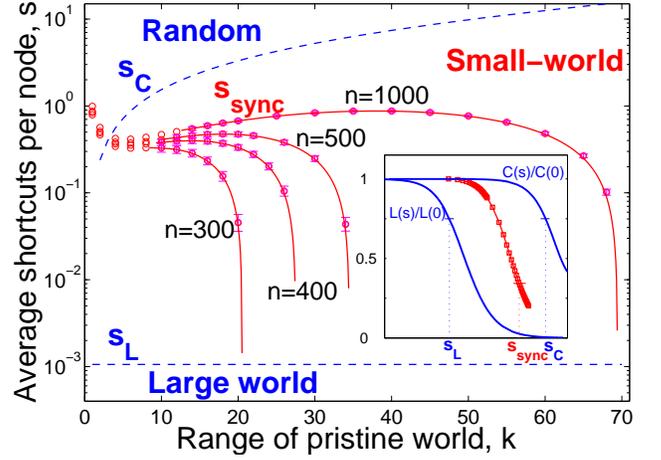}
\caption{Synchronizability thresholds $s_{\rm{sync}}(\circ)$ for
graphs with $n$ nodes ($n=300, 400, 500, 1000$) and range $k \in
[1,70]$, numerically averaged over 1000 realizations. The solid
lines correspond to the analytical Eq.~(\ref{eq:analyticapprox}),
valid in the range $n^{1/3} < k < k_{\min}(n)$. For most
parameters, $s_{\rm{sync}}$ lies within the small-world region
between the dashed lines ($ s_L < s < s_C $) depicted here for
$n=1000$, but it is distinct from the SW onset $s_L$. Note how
synchronization is achievable without random shortcuts by
increasing the deterministic range up to $k_{\rm{min}}(n)$ (see
Fig.~{\ref{fig:semirandom}}). Inset: decay of the average distance
$L$, clustering $C$, and eigenratio (squares) as shortcuts are
added to a pristine world of range $k=20$ and $n=500$. We define
$s_L$ and $s_C$ as the points where $L$ and $C$ are 75\% of the
pristine world value; $s_{\rm{sync}}$ is the point where the
eigenratio $\theta_{\rm{max}}/\theta_1 = \beta \equiv 37.85$. }
\label{fig:threshold}
\end{figure}

The numerical results in Fig.~\ref{fig:threshold} illustrate the
SW effect on the synchronization of networks of different size and
range. For concreteness, all our numerics have been performed for
a network of identical $x$-coupled R\"ossler chaotic oscillators
with $\beta \simeq 37.85$. Similarly to other locally connected
networks, pristine worlds have a large eigenratio
$\theta_{\rm{max}}/\theta_1$ (i.e., they are difficult to
synchronize). However, as $s$ is increased the eigenratio falls
sharply until, at a value $s_{\rm{sync}}$, the
condition~(\ref{eq:critcond}) is reached (i.e., the addition of
shortcuts makes it synchronizable). The dependence of
$s_{\rm{sync}}$ on the network parameters $\{n,k\}$ is notably
complicated. First, there appears to be an optimal range $k \simeq
4$ for which the SW is most efficient. Moreover, the
synchronization threshold $s_{\rm{sync}}$ lies in the small-world
region but does not seem to coincide with its onset. The SW onset
($s_L$) is defined~\cite{refWS,refDuncanbook} by the decay of the
average graph distance~\cite{refNMW} $L(s) \simeq (n/k) \, f(n
s)$, where $f(x)=\arg \tanh \left (x/\sqrt{x^2+2x} \right ) /
\sqrt{4(x^2+2 x)}$. Fixing $L(s_L)/L(0) = 3/4$, we obtain $s_L
\simeq 1.061/n, \: n k \gg 1$. The end of the SW region ($s_C$)
corresponds to the effective graph randomization through the loss
of transitivity~\cite{refWS,refNSW,refDuncanbook}, as given by the
decay of the clustering coefficient
$C$~\cite{refBW,refMarkpersonal}: $C(s)/C(0) \simeq (2k-1)/(2k
(1+s/k)^2 -1), \: n \gg 1$. Again, fixing $C(s_C)/C(0) = 3/4$, we
obtain $s_C \simeq k \left (-1+ [(8k-1)/(6k-3)]^{-1/2} \right )
\simeq 0.155 \; k$. The synchronization threshold generally lies
between these two boundaries which scale differently with $n$ and
$k$: $s_L \simeq 1.061/n < s_{\rm{sync}} < s_{C} \simeq 0.155 \;
k$.

We can gain insight into the synchronization threshold through an
analytical perturbation of the eigenratio of the SW Laplacian $G=
G^0 + G^r$. Here, $G^0$ is the deterministic Laplacian of the
pristine world, and $G^r$ is the stochastic Laplacian for the
random shortcuts: $G^r_{ij}=G^r_{ji}=-\xi_{ij}$ (for $i+k+1 \leq j
\leq \min \{n,n-k+i-1\}$ with $1 \leq i \leq n-k+1$); $G^r_{ij}=0$
(otherwise); and $G^r_{ii}=-\sum_{j=1}^n G^r_{ij}$ (for $1 \leq i
\leq n$). The $\xi_{ij}$ are $n(n-2k-1)/2 \,$ i.i.d. Bernoulli
random variables which take the value 1 with probability $q/n
\equiv 2 s/(n-2k-1)$ (and the value 0 with probability $1-q/n$).
The circulant $G^0$ is Fourier-diagonalizable~\cite{refRM} with
spectrum $\theta^0_j = (2k+1)- \sin [(2k+1) \pi j/n] /\sin [\pi
j/n], \,\, 1 \leq j \leq n-1$, (plus $\theta^0_0=0$ of any
Laplacian). The FNZE and the maximum eigenvalue of the unperturbed
lattice are:
\begin{eqnarray}
\theta^0_1 &\simeq& 2 \pi^2 k (k+1) (2k+1)/(3 n^2), \;\; k \ll n
\label{eq:fnze0}\\
\theta^0_{\rm{max}} &\simeq& (2k+1) +
\csc \left[\frac{3 \pi/2}{2k+1} \right] \label{eq:emax0_1} \\
&\simeq& (2k+1) [1+2/3 \pi], \;\; k \gg 1, \label{eq:emax0}
\end{eqnarray}
where (\ref{eq:emax0_1}) follows from a continuum approximation.

Following an ``honest'' treatment~\cite{refBoyce} with $G^r$ as
the perturbation, we treat the analytical expressions of the
doubly degenerate eigenvalues as random variables to obtain their
expectations. We postpone the detailed
calculations~\cite{refuslong} and sketch here the main results.
After some stochastic calculus, the expectations of the
eigenvalues of the SW Laplacian to second order are shown to be:
\begin{eqnarray}
{\mathcal E}\theta^{(1)}_i &\simeq& q \pm \sqrt{3 \pi q/4 n}
\label{eq:expectation1}\\
{\mathcal E}\theta^{(2)}_i &\simeq& \frac{2 q}{n}
{{\sum_{m=1}^n}^\prime (\theta^0_i-\theta^0_m})^{-1},
\label{eq:expectation2}
\end{eqnarray}
for $q/n$ and $k/n$ small. To improve the accuracy of
$s_{\rm{sync}}$, we have also obtained an approximation
to~(\ref{eq:expectation2}) for FNZE:
\begin{eqnarray}
{\mathcal E}\theta^{(2)}_1 \simeq \frac{-2 q}{K^3} \left [ \frac{9
n}{\pi^2} + K^2 - \left( \frac{7}{5}+\frac{6}{\pi^2}\right) K -
\frac{2}{\pi} \right ] , \label{eq:approx22}
\end{eqnarray}
where $K=2 k +1$. Eqns.
(\ref{eq:critcond}),(\ref{eq:fnze0}),(\ref{eq:emax0}),(\ref{eq:expectation1}),
and (\ref{eq:approx22}) are then used to obtain an estimate of
$s_{\rm{sync}}$ as the solution of an algebraic equation involving
only $n$ and $k$:
\begin{equation}
\theta^0_{\rm{max}} + {\mathcal E}\theta^{(1)}_{\rm{max}} = \beta
\left( \theta^0_1 +{\mathcal E}\theta^{(1)}_1 + {\mathcal
E}\theta^{(2)}_1 \right). \label{eq:analyticapprox}
\end{equation}
As shown in Fig.~\ref{fig:threshold}, this approximates well our
numerics for $n^{1/3} < k \ll n$,  where the
Rayleigh-Schr\"odinger perturbation expansion is valid.

Using~(\ref{eq:analyticapprox}), we can obtain~\cite{refuslong} a
first order estimate for the \textit{maximum} of the
synchronization threshold $s_{\rm{sync}}^*$ in the valid range.
The maximum occurs at $k^* \simeq n \sqrt{(1+2/3 \pi)/2 \pi^2
\beta}$ with the asymptotic value $s_{\rm{sync}}^* \simeq (2+4/3
\pi) (1-2 k^*/n)/3 (\beta -1)$. Therefore, $s_{\rm{sync}} < s_C \;
[2 (1+2/3 \pi)/(2 \surd{3} -3) (\beta -1)]$ is bounded by the end
of the SW region but linked to it. For $k<n^{1/3}$, the eigenvalue
bunching (quasi-degeneracy) in the pristine lattice renders the
doubly degenerate perturbation invalid. We are developing another
approximation to quantify the behavior in this limit, but our
numerics~\cite{refuslong} indicate that the dependence of
$s_{\rm{sync}}$ with $n$ is sub-logarithmic. This confirms that
the synchronizability is most effectively improved for small-range
networks (Fig.~\ref{fig:threshold}).

\begin{figure}
\includegraphics[width=3.4in]{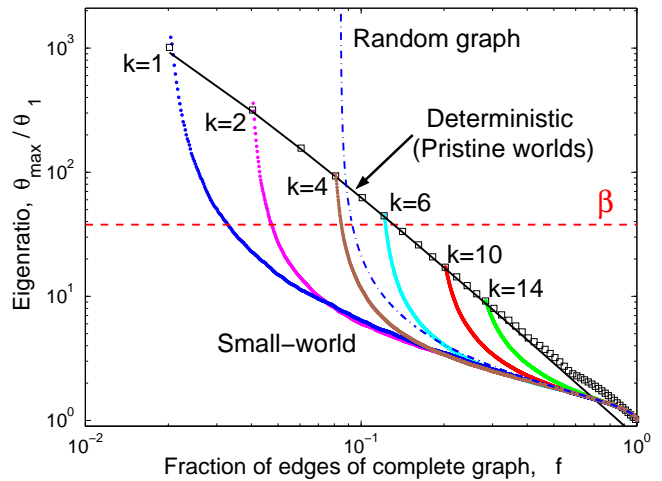}
\caption{Eigenratio decay in a $n=100$ lattice as $f$ edges are
added following purely deterministic, semi-random (SW), and purely
random schemes. Networks become synchronizable below the dashed
horizontal line ($\beta$). The squares (numerical) and the solid
line (Eq.~(\ref{eq:synchropristine})) show the decrease of the
eigenratio of pristine worlds ($k$-cycles) through the
deterministic addition of short-range connections---for $n=100$,
networks with $k \geq 7$ are synchronizable. The semi-random SW
approach (dots, shown for $k=1,2,4,6,10,14$) is more efficient in
producing synchronization. The dot-dashed line corresponds to
purely random graphs (RG, Eq.~(\ref{eq:synchrorandom})), which
become \textit{almost surely} disconnected at $f \simeq 2 \ln
n/(n+2 \ln n) = 0.0843$ (thus, with $\theta^{RG}_1=0$ and
unsynchronizable). The merging of the SW and RG behaviors as $f
\to 1$ is the dynamical analogue of the effective randomization
that leads to $s_C$.} \label{fig:semirandom}
\end{figure}

How efficient is the addition of random shortcuts for
synchronization? We have compared the semi-random SW approach with
purely random and purely deterministic schemes. An example of the
latter is the synchronization of pristine worlds through the
increase of the range $k$. From (\ref{eq:fnze0}) and
(\ref{eq:emax0}), the eigenratio of a pristine lattice is
\begin{equation}
\frac{\theta^0_{\rm{max}}}{\theta^0_{1}}
\simeq \frac{3 \pi+2}{2 \pi^3} \, \frac {n^2}{ k (k+1)}.
\label{eq:synchropristine}
\end{equation}
Therefore, $n$ nodes can be synchronized in a $k$-cycle if $k
> k_{\rm{min}} \simeq n \sqrt{(3 \pi +2)/2 \pi^3 \beta}$.
(Note in Fig.~\ref{fig:threshold} the consistency of our
analytical approximation (\ref{eq:analyticapprox}):
$s_{\rm{sync}}=0$ precisely at $k_{\rm{min}}$.) For purely random
graphs ${\mathcal G}_{n,f}$~\cite{refmohar},
\begin{equation}
\frac{\theta^{RG}_{\rm{max}}}{\theta^{RG}_1} \simeq
\frac{nf-\sqrt{2f(1-f)n \ln n}}{nf+\sqrt{2f(1-f)n \ln n}}\, ,
\label{eq:synchrorandom}
\end{equation}
where $f$ is the number of edges measured as a fraction of the
complete graph. These are compared with SW graphs in
Fig.~\ref{fig:semirandom}.

We remark on several observations regarding
Fig.~\ref{fig:semirandom}. First, the SW addition of shortcuts is
more efficient than the deterministic addition of short-range
layers. Second, the effective randomization of SW lattices with
edge addition translates into converging synchronization behaviors
of random and SW networks at large $f$. The $f \to 1$ region is
thus \textit{robustly} stable: cutting connections from the fully
connected graph has very little effect on synchronization
stability---not until over 90\% are cut (for $k$ small) does the
eigenratio begin to change drastically. Moreover, if we interpret
the number of edges needed to synchronize $n$ nodes as a simple
measure of ``cost'', adding many connections buys little extra
stability beyond the small-world regime. Finally, it can be
shown~\cite{refuslong} that the general trends of the eigenratio
in Fig.~\ref{fig:semirandom} (namely, ``hyperbolic'' dependence
for $f$ small, and near independence for $f$ large) can be
predicted with the naive perturbation result that both $\theta_1$
and $\theta_{\rm max}$ change linearly with the number of added
connections.

\begin{figure}
\includegraphics[width=3.4in]{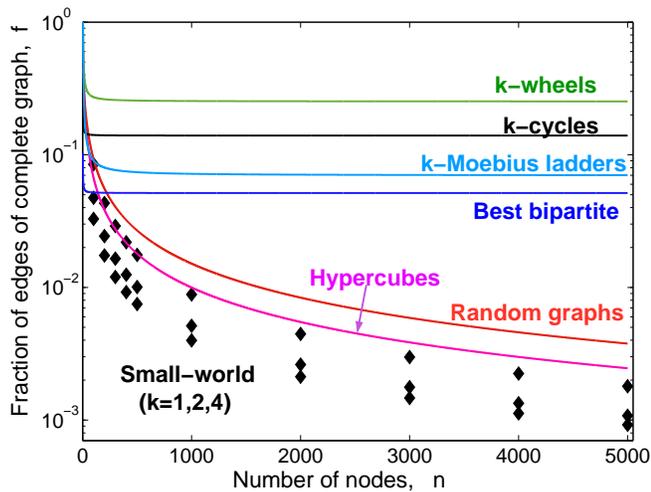}
\caption{``Cost'' of synchronization measured as the number of
edges needed to synchronize a lattice of $n$ chaotic R\"ossler
systems arranged in different topologies: deterministic graphs
($k$-wheels, $k$-cycles, $k$-M\"obius ladders, bipartites,
hypercubes), random graphs, and small-worlds ($\blacklozenge$).
Small-worlds scale favorably compared to deterministic structures
(and comparably to the ideal and largely unrealizable hypercubes).
Also, SW graphs with small range $k$ are as cost-efficient as
random graphs but demand less (algorithmic) storage memory.}
\label{fig:design}
\end{figure}

We have also compared the synchronization cost (in edges) for SW
systems and regular (constructive) lattices
(Fig.~\ref{fig:design}). For $x$-R\"ossler systems in a $k$-cycle,
this cost tends to $f=2 k_{\rm{min}}/(n-1) \simeq 0.140$. Other
constructive lattices~\cite{refSchwenk,refuslong} also tend to
constant fractions: $f=0.252$ (for $k$-wheels), $f=0.070$ (for
$k$-M\"obius ladders), and $f=0.053$ (for the most economical
bipartite graph). In all those cases, the cost of synchronization
is high: the necessary number of edges scales like $\sim n^2$,
just like the complete graph. At the other end of deterministic
graphs lies the quasi-optimal (though virtually unrealizable)
hypercube, which is always synchronizable with a number of edges
$f \sim \log_2 n /n$. This behavior is similar to that of random
graphs: from Eq.~(\ref{eq:synchrorandom}) {\it almost sure}
synchronization of ${\mathcal G}_{n,f}$ is asymptotically achieved
when $f \sim \ln n/n$. Remarkably, Fig.~\ref{fig:design} shows
that the cost of synchronizing small-$k$ SW networks is low, i.e.,
comparable to that of random graphs and
hypercubes~\cite{refChungmatching}.

These results hint at research that could deepen the connections
between topology and dynamics on networks. With a view to improved
design, the dynamic eigenratio criterion can be related to other
graph-theoretical properties (e.g., connectivity, diameter, and
convergence of Markov chains)~\cite{refChung}. Moreover, other
measures of cost (e.g., robustness under edge deletion) should be
considered as possible design constraints. Finally, recent results
could lead to extensions of this work to incorporate more general
concepts of stability~\cite{refpablo}, and broader definitions of
small-world lattices~\cite{refNSW}.

We thank Steve Strogatz for his deep and insightful involvement in
this work, and Mark Newman for sharing computer code and
unpublished results.

\bibliography{SW4}

\end{document}